\documentclass[12pt, letterpaper]{article}

\usepackage{amsmath, amssymb}%Needed for multline and symbols
\usepackage{cite}%Needed for long citations
\usepackage{fancyhdr}%Needed for title page headings
\usepackage[top=1in, bottom=1.5in, left=0.9in, right=0.9in]{geometry}%Needed for margins
\usepackage[dvipdfm, dvips]{graphicx}%Needed for includegraphics
\usepackage{hyperref}%Needed for hyper references

\numberwithin{equation}{section}%Needed for equation numbers to follow sections
\begin{document}

%%%%%%%%%%%%%%%%%%%%%%%%%%%%%%%%%%%%%%%%%%%%%%%%%%%%%%%%%%%%%%%%%%%%%%%%%%%%%%%%%%%%%%%%%%%%%%%%%%%%%%%%%%%%%%%%%%%%%%%%%%%%%%%%%%%%%%%%%%%%%%%%%%%%
%%%%%%%%%%%%%%%%%%%%%%%%%%%%%%%%%%%%%%%%%%%%%%%%%%%%%%%%%%%%%%%%%%%%%%%%%%%%%%%%%%%%%%%%%%%%%%%%%%%%%%%%%%%%%%%%%%%%%%%%%%%%%%%%%%%%%%%%%%%%%%%%%%%%

\setcounter{page}{0}%Needed for the title page to have (virtual) number 0
\date{}%Needed to remove the date

\lhead{}\chead{}\rhead{\footnotesize{RUNHETC-2010-19\\SCIPP-10-14\\UCSD-PTH-10-06}}\lfoot{}\cfoot{}\rfoot{}

\title{\textbf{Direct Detection of Dark Matter\vspace{0.15cm}\\
Electromagnetic Dipole Moments\vspace{0.7cm}}}

\author{Tom Banks$^{1,2}$, Jean-Fran\c{c}ois Fortin$^3$ and Scott Thomas$^1$\vspace{0.7cm}\\
{\normalsize{$^1$NHETC and Department of Physics and Astronomy, Rutgers University,}}\\
{\normalsize{Piscataway, NJ 08854-8019, USA}}\vspace{0.2cm}\\
{\normalsize{$^2$SCIPP and Department of Physics, University of California, Santa Cruz}}\\
{\normalsize{Santa Cruz, CA 95064-1077, USA}}\vspace{0.2cm}\\
{\normalsize{$^3$Department of Physics, University of California, San Diego}}\\
{\normalsize{La Jolla, CA 92093-0319, USA}}}

\maketitle
\thispagestyle{fancy}%Needed to remove the page number from the title page

\begin{abstract}
\normalsize
\noindent
Dark matter candidates with electromagnetic dipole moments can arise as dark baryons in gauge-mediated or technicolor models.  These dark matter candidates interact with nuclei in direct detection experiments mainly through magnetic and/or electric dipole moments.  The scattering cross sections depend on the nuclear magnetic moments and nuclear charge and have an infrared enhancement compared with typical WIMP constant contact interactions, leading to distinctive nuclear recoil energy spectra.  These characteristics result in an enhanced signal for the DAMA experiment compared with the CDMS or XENON experiments.  The positive results of DAMA, along with the null results of CDMS and XENON, are consistent with a dark matter particle with magnetic dipole moment and a mass around ten GeV.  Significant direct detection signals can arise from dipolar dark matter with mass up to of order tens of TeV.
\end{abstract}

%%%%%%%%%%%%%%%%%%%%%%%%%%%%%%%%%%%%%%%%%%%%%%%%%%%%%%%%%%%%%%%%%%%%%%%%%%%%%%%%%%%%%%%%%%%%%%%%%%%%%%%%%%%%%%%%%%%%%%%%%%%%%%%%%%%%%%%%%%%%%%%%%%%%

\newpage
\tableofcontents
\vspace{1cm}

%%%%%%%%%%%%%%%%%%%%%%%%%%%%%%%%%%%%%%%%%%%%%%%%%%%%%%%%%%%%%%%%%%%%%%%%%%%%%%%%%%%%%%%%%%%%%%%%%%%%%%%%%%%%%%%%%%%%%%%%%%%%%%%%%%%%%%%%%%%%%%%%%%%%

\section{Introduction}\label{sec:intro}

The two most popular candidates for particle dark matter are a WIMP (with particular emphasis on the LSP in gravity-mediated SUSY breaking models) and the QCD axion.  However, in a wide range of technicolor and hidden sector gauge mediation models, the most natural dark matter candidate is a neutral particle carrying an approximately conserved baryon number like quantum number in the hidden sector \cite{Nussinov:1985xr,Chivukula:1989qb,Chivukula:1992pn,Bagnasco:1993st,Banks:2005hc}.  The relic abundance of this particle is not thermal, but is determined by a primordial asymmetry generated in the very early universe.  Such models do not have the ``WIMP miracle'', but do open up the possibility of explaining the ratio between baryon and dark matter densities \cite{Kaplan:1991ah}.

We have two primary motivations for interest in these candidates.  Gauge-mediated SUSY breaking resolves many of the otherwise puzzling flavor (and perhaps CP) problems of gravity mediation, but does not usually have a natural WIMP (see however, \cite{Feng:2008ya}).  Axion dark matter is also potentially problematic in gauge mediation \cite{Carpenter:2009sw}, especially for direct mediation models of the kind studied recently by two of the authors \cite{Banks:2009rb}.  It is difficult to get the superpartner of the axion to decay in a way that does not spoil nucleosynthesis.

A much more pragmatic reason for interest in these \textit{dark baryons} is that they will generically have magnetic, and perhaps electric, dipole moments.  As a consequence, their signals in terrestrial dark matter detectors are \textit{very different from those of conventional WIMPs}.  The infrared tail of dipole charge scattering can dramatically change the cross sections for very low recoil energies.  As we will see, dark baryons can give low energy signals even for very large masses.  In addition, some of the dipole scattering cross sections are very different for different nuclear targets, and lead to different conclusions in comparing the implications of one experiment for another.  We will, for example, show that there are particular dark baryon parameters that are compatible with {\it all} claims about direct detection experiments, including DAMA.  More generally, we find that it is typical to have larger rates in DAMA than in other experiments, because of the properties of iodine, but not always enough to account for the discrepancy between DAMA and CDMS/XENON.  Another interesting feature of the infrared tail of dipole scattering is that it allows for rather large masses of the dark baryons, consistent with experimental observations.  Since most direct mediation or technicolor models will have dark baryon masses in the range of a few GeV to a few tens of TeV, this remark is quite significant.

This paper is an update of \cite{Bagnasco:1993st}, which includes predictions for all current and planned dark matter direct detection experiments.  We have carefully redone the calculations to determine whether disagreements with other papers in the literature were significant, and we verify the results of \cite{Bagnasco:1993st} are correct.

The plan of this paper is as follows : In Section \ref{sec:DMDM} we quickly review the theory of fermions with magnetic and electric dipole moments.  We also include the appropriate differential cross sections with nuclei and compare them with the usual differential cross section obtained from a WIMP with constant contact interactions.  Section \ref{sec:Exp} discusses the current and planned dark matter direct detection experiments and summarizes the relevant features of each experiment.  Sections \ref{subsec:CDMSXENON} and \ref{subsec:DAMA} focus on CDMS/XENON and DAMA respectively.  CDMS and XENON contour plots for the magnetic and electric dark matter Land\'{e} factors are shown in function of the dark matter mass for several exposures.  DAMA plots for the magnetic and electric dipole moment modulation amplitudes are shown as functions of the nuclear recoil energy, and are compared to DAMA data.  Finally we discuss the implications of fermionic dark matter with dipole moments and conclude in Section \ref{sec:conc}.  Physical parameters and computations are shown in Appendix \ref{app:EffLag}.

%%%%%%%%%%%%%%%%%%%%%%%%%%%%%%%%%%%%%%%%%%%%%%%%%%%%%%%%%%%%%%%%%%%%%%%%%%%%%%%%%%%%%%%%%%%%%%%%%%%%%%%%%%%%%%%%%%%%%%%%%%%%%%%%%%%%%%%%%%%%%%%%%%%%

\section{Dark Matter with Dipole Moments}\label{sec:DMDM}

The magnetic and electric dipole moment operators for fermionic dark matter (DM) are the only dimension $5$ operators coupling to the SM and thus are the most relevant operators for fermionic DM direct detection experiments.  For a fermionic DM $\psi$ of mass $m_{\rm DM}$, the effective Lagrangian is \cite{Bagnasco:1993st}
\begin{equation}\label{eqn:EffLag}
\delta\mathcal{L}_{\rm DM}=\bar{\psi}(i\gamma^\mu\partial_\mu-m_{\rm DM})\psi+\frac{g_Me}{8m_{\rm DM}}\bar{\psi}\sigma^{\mu\nu}\psi F_{\mu\nu}+\frac{g_Ee}{8m_{\rm DM}}\bar{\psi}\sigma^{\mu\nu}\psi \widetilde{F}_{\mu\nu}
\end{equation}
where the fermionic DM magnetic and electric gyromagnetic ratios are $g_M\sim4m_{\rm DM}/\Lambda_{\rm DM}$ and $g_E\sim4m_{\rm DM}\Lambda_{\rm DM}/\Lambda_{\rm CP}^2$ respectively with $\Lambda_{\rm DM}$ the dark sector scale and $\Lambda_{\rm CP}>\Lambda_{\rm DM}$ the CP-violating scale.  Due to the intrinsic P and T violation of the fermionic DM electric dipole moment operator, the electric dipole moment operator must be suppressed in any model that is consistent with experimental constraints on CP violation.  We have parametrized this suppression by writing the small $g_E$ in terms of a ratio of scales.  We emphasize that there might be models in which CP violation is suppressed instead by powers of a dimensionless coupling.  The scale $\Lambda_{CP}$ would not then be a threshold for new physics.

\begin{table}[t]
\begin{center}
\includegraphics[scale=0.85]{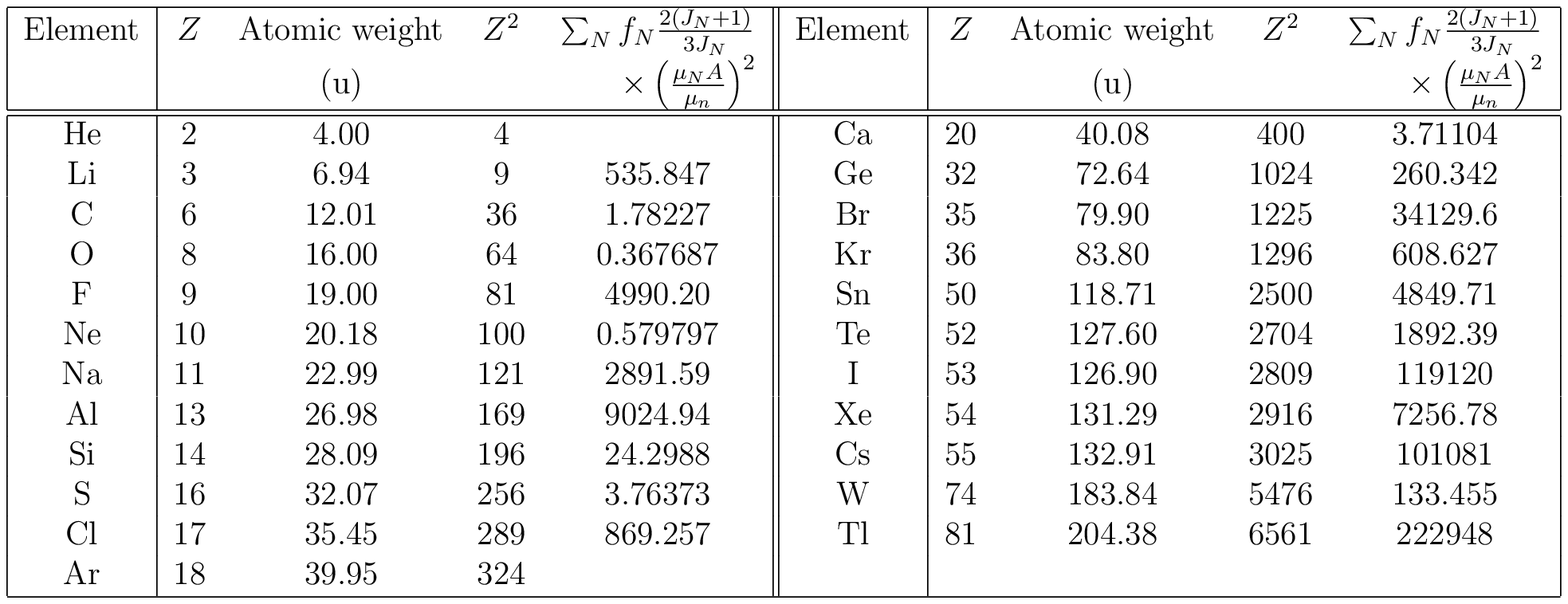}
\caption{Elements used in DM direct detection searches with related features \cite{Lide:2009}.  For each element, the last two columns give the dipole moment differential cross sections relative ``strength'' for fermionic DM-nucleus scatterings as dictated by equations (\ref{eqn:CrossM}) and (\ref{eqn:CrossE}) and table \ref{tab:Isotopes}.}
\label{tab:Elements}
\end{center}
\end{table}

The Lagrangian equation (\ref{eqn:EffLag}) gives the fermionic DM a magnetic dipole moment $\mu_{\rm DM}=g_Me/4m_{\rm DM}$ and an electric dipole moment $d_{\rm DM}=g_Ee/4m_{\rm DM}$ consistent with the usual definitions.  In strongly-coupled theories the fermionic DM Land\'{e} factors $g_M$ and $(\Lambda_{\rm CP}/\Lambda_{\rm DM})^2g_E$ should be numbers of order one while in weakly-coupled theories they should be numbers suppressed by a loop factor.

For direct detection experiments the magnetic and electric dipole moment differential cross sections in the non-relativistic limit in the lab frame
are given by (see Appendix \ref{app:EffLag})
\begin{eqnarray}
\frac{d\sigma_M}{dE_R} &=& \frac{\pi(g_M\alpha)^2}{4(m_{\rm DM}+m_N)^2E_R^{\rm max}}\left[\frac{2(J+1)}{3J}\left(\frac{\mu_NA}{\mu_n}\right)^2|F_s(E_R)|^2\right.\nonumber\\
 && \hspace{1cm}\left.+Z^2\left(\frac{(m_{\rm DM}+m_N)^2}{m_{\rm DM}^2}\frac{E_R^{\rm max}}{E_R}-\frac{2m_N}{m_{\rm DM}}-1\right)|F_c(E_R)|^2\right]\label{eqn:CrossM}\\
\frac{d\sigma_E}{d E_R} &=& \frac{\pi(g_EZ\alpha)^2}{2(m_{\rm DM}+m_N)^2E_R^{\rm max}}\frac{m_N}{E_R}|F_c(E_R)|^2\label{eqn:CrossE}.
\end{eqnarray}
Here $m_N$ is the nucleus mass on which the fermionic DM recoils, $Z$ and $A$ are the atomic and mass numbers of the nucleus and $J$ and $\mu_N/\mu_n$ are the nuclear spin and nuclear magnetic moment respectively.  Moreover $E_R$ is the nuclear recoil kinetic energy in the lab frame and $E_R^{\rm max}=2m_Nm_{\rm DM}^2v^2/(m_{\rm DM}+m_N)^2$ is the maximum nuclear recoil energy, $E_R\leq E_R^{\rm max}$.  Finally $F_s(E_R)$ and $F_c(E_R)$ are the appropriate form factors (see Appendix \ref{app:EffLag} for more details).  For a dark matter particle of general spin $J_\psi$ the cross sections above are multiplied by $\frac{4}{3}J_\psi(J_\psi+1)$.

Notice that the magnetic dipole moment differential cross section equation (\ref{eqn:CrossM}) has a spin-independent (SI) and a spin-dependent (SD) contribution.  The latter comes from the coherent coupling of the magnetic dipole moment with the nuclear charge current in the fermionic DM rest frame.  Notice moreover that the SI contribution to the magnetic dipole moment differential cross section as well as the electric dipole moment differential cross section have an infrared (IR) enhancement.  This is in contrast with the usual constant contact interaction differential cross section for WIMPs in the non-relativistic limit which is \textit{not} enhanced in the IR.  Indeed, the differential cross section for s-wave nucleon scattering in the non-relativistic limit is (see Appendix \ref{app:EffLag})
\begin{equation}\label{eqn:CrossCI}
\frac{d\sigma_{CI}}{d E_R}=\frac{A^2\sigma_n}{E_R^{\rm max}}\frac{\widetilde{m}_{{\rm DM},N}^2}{\widetilde{m}_{{\rm DM},n}^2}|F_c(E_R)|^2.
\end{equation}
Here $\sigma_n\equiv\sigma({\rm DM}+n\rightarrow{\rm DM}+n)$ is the total cross section per nucleon while $\widetilde{m}_{{\rm DM},N}$ and $\widetilde{m}_{{\rm DM},n}$ are the reduced masses of the fermionic DM-nucleus and fermionic DM-nucleon system respectively.  This difference significantly changes the cross section for low nuclear recoil energies.

\begin{table}[t]
\begin{center}
\includegraphics[scale=0.85]{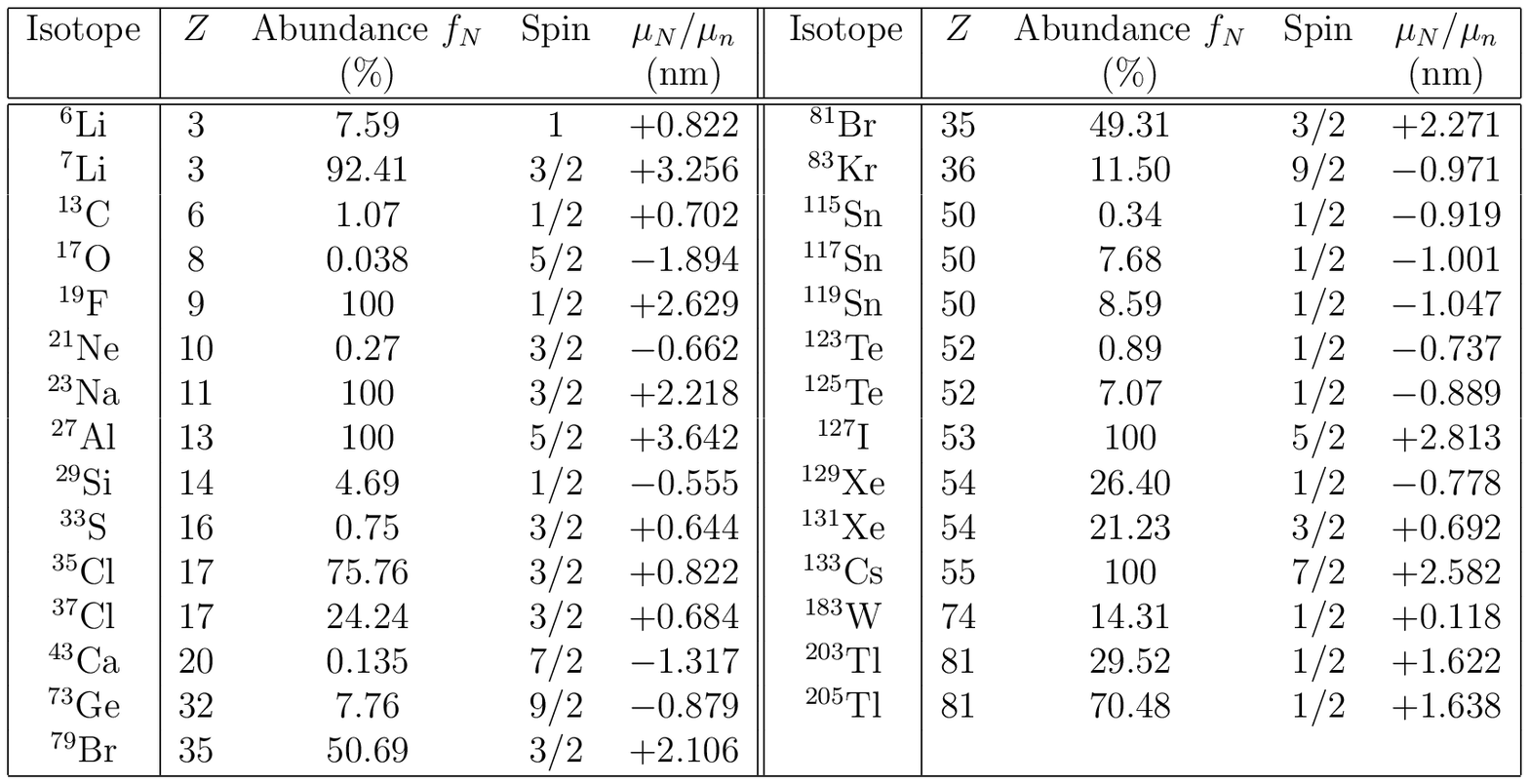}
\caption{Natural (stable) isotopes with non-vanishing nuclear magnetic moments relevant in DM direct detection searches with related features \cite{Lide:2009}.}
\label{tab:Isotopes}
\end{center}
\end{table}

Finally, in order to compare theory with experiment, the total number of events seen for each experiment must be computed from the differential scattering rate per unit detector mass equation (\ref{eqn:DiffScatRate}).  All other relevant physical parameters and computations are discussed in Appendix \ref{app:EffLag}.  For example the form factors is given by equation (\ref{eqn:FcFs}) and the fermionic DM speed distribution in the lab frame is given by equation (\ref{eqn:DMfv}).

Before focusing on the experiments, we would like to emphasize that the DM energy loss $\Delta E$, due to elastic scatterings with the constituents of the earth's atmosphere and the earth's crust,
\begin{equation}
\Delta E=-n_NL\int dE_R\,E_R\frac{d\sigma}{dE_R},
\end{equation}
is completely negligible for generic gyromagnetic ratios of the size we assume, contrary to the claim of\cite{Sigurdson:2004zp}.  Here $n_N$ is the number of atoms per unit volume in the earth's atmosphere and earth's crust (which we approximate by silicon only), $L$ is the distance traveled in the earth's atmosphere and the earth's crust by the fermionic DM before reaching the experiment and $d\sigma/dE_R$ is the differential cross section.  The DM loss of energy is several orders of magnitude smaller than the DM initial kinetic energy in the lab rest frame and is therefore unimportant.

%%%%%%%%%%%%%%%%%%%%%%%%%%%%%%%%%%%%%%%%%%%%%%%%%%%%%%%%%%%%%%%%%%%%%%%%%%%%%%%%%%%%%%%%%%%%%%%%%%%%%%%%%%%%%%%%%%%%%%%%%%%%%%%%%%%%%%%%%%%%%%%%%%%%

\section{Direct Detection Experiments}\label{sec:Exp}

\begin{table}[t]
\begin{center}
\includegraphics[scale=0.85]{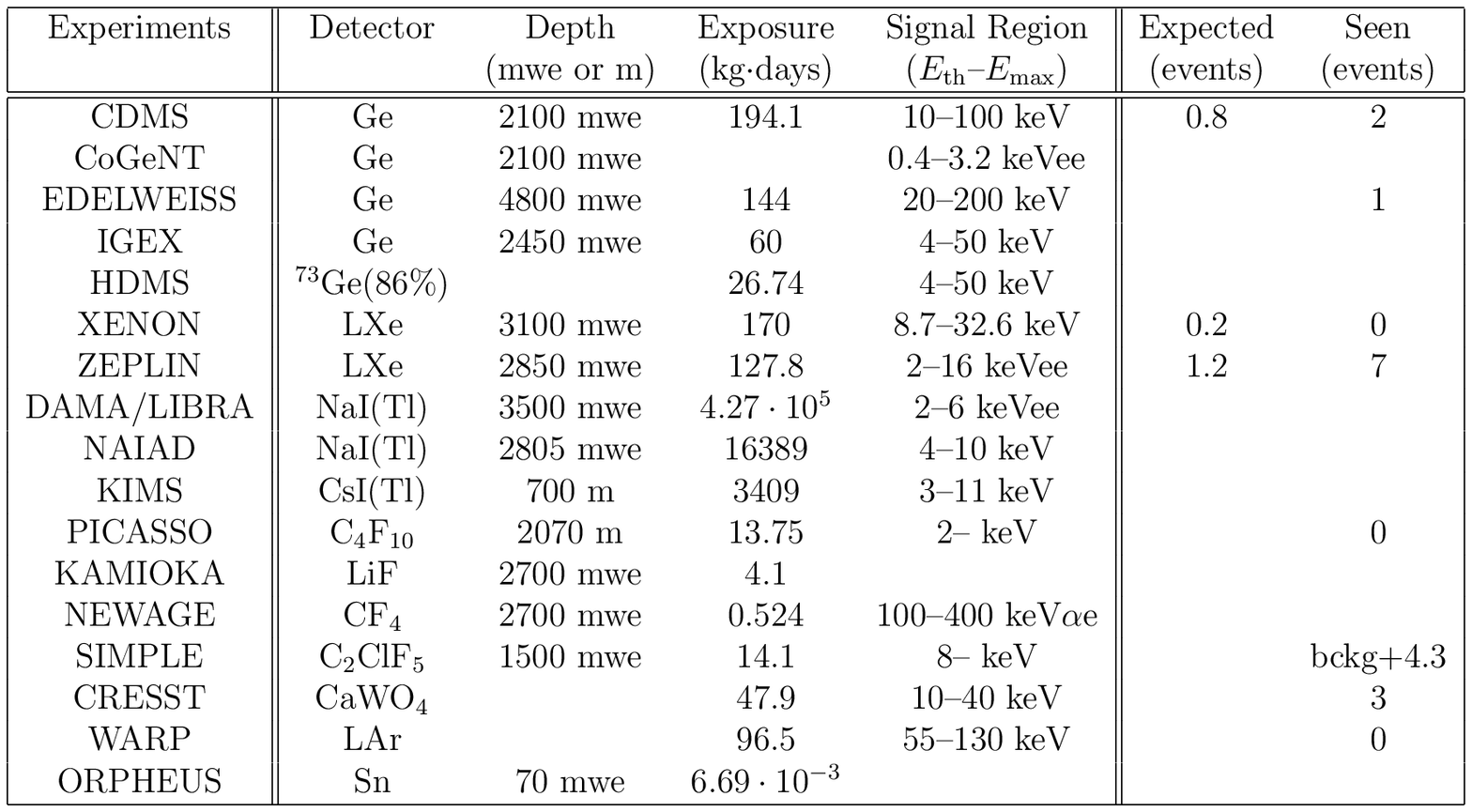}
\caption{Relevant direct search experiments with related features.}
\label{tab:Exp}
\end{center}
\end{table}

Direct detection experiments hope to observe DM directly by its elastic scattering with the nuclei present in the detectors.  The most common elements used in the direct detection experiment detectors are listed in table \ref{tab:Elements} with their relevant features.  For each element, the first column gives the atomic number of the nucleus, the second column provides the atomic weight of the nucleus and the third and fourth columns supply the dipole moment differential cross sections relative ``strength'' for the SI magnetic and electric dipole moments and for the SD magnetic dipole moment respectively as can be seen from equations (\ref{eqn:CrossM}) and (\ref{eqn:CrossE}).  The fourth column is computed from table \ref{tab:Isotopes} which enumerates all stable isotopes with non-vanishing nuclear magnetic moments.

\begin{figure}[p]
\begin{minipage}[b]{0.5\linewidth}
\begin{center}
\includegraphics[scale=0.3]{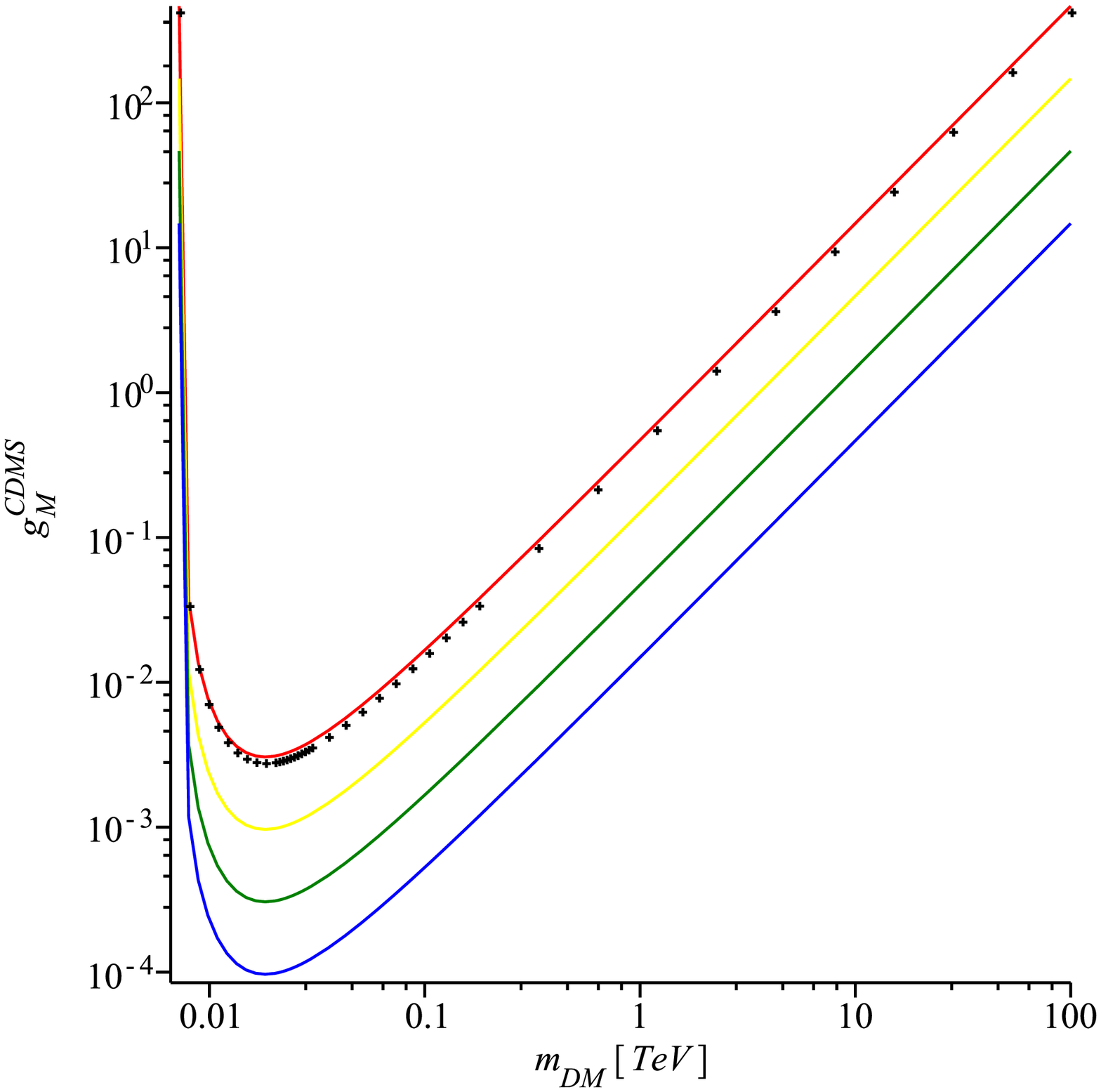}
\includegraphics[scale=0.3]{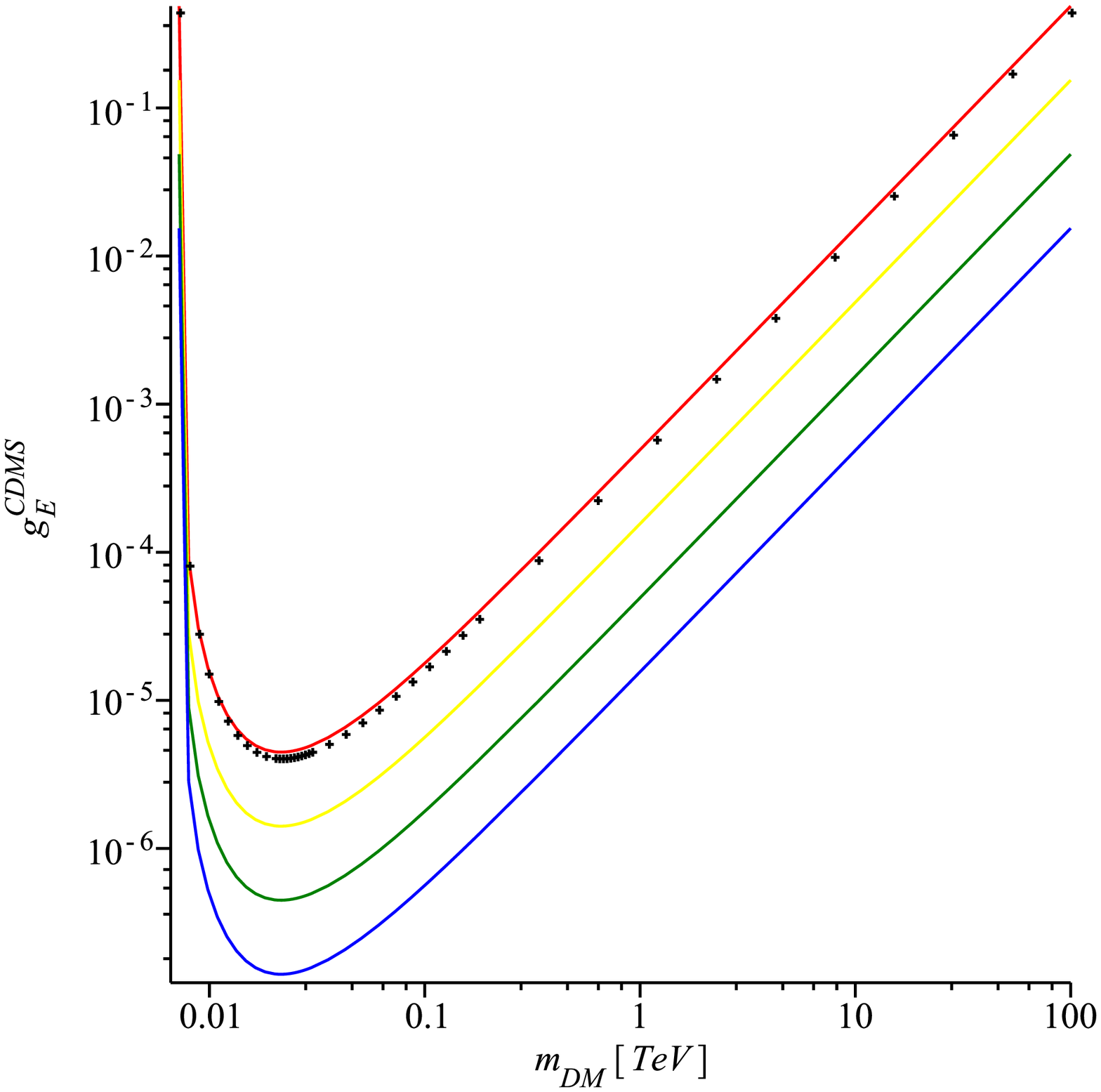}
\includegraphics[scale=0.3]{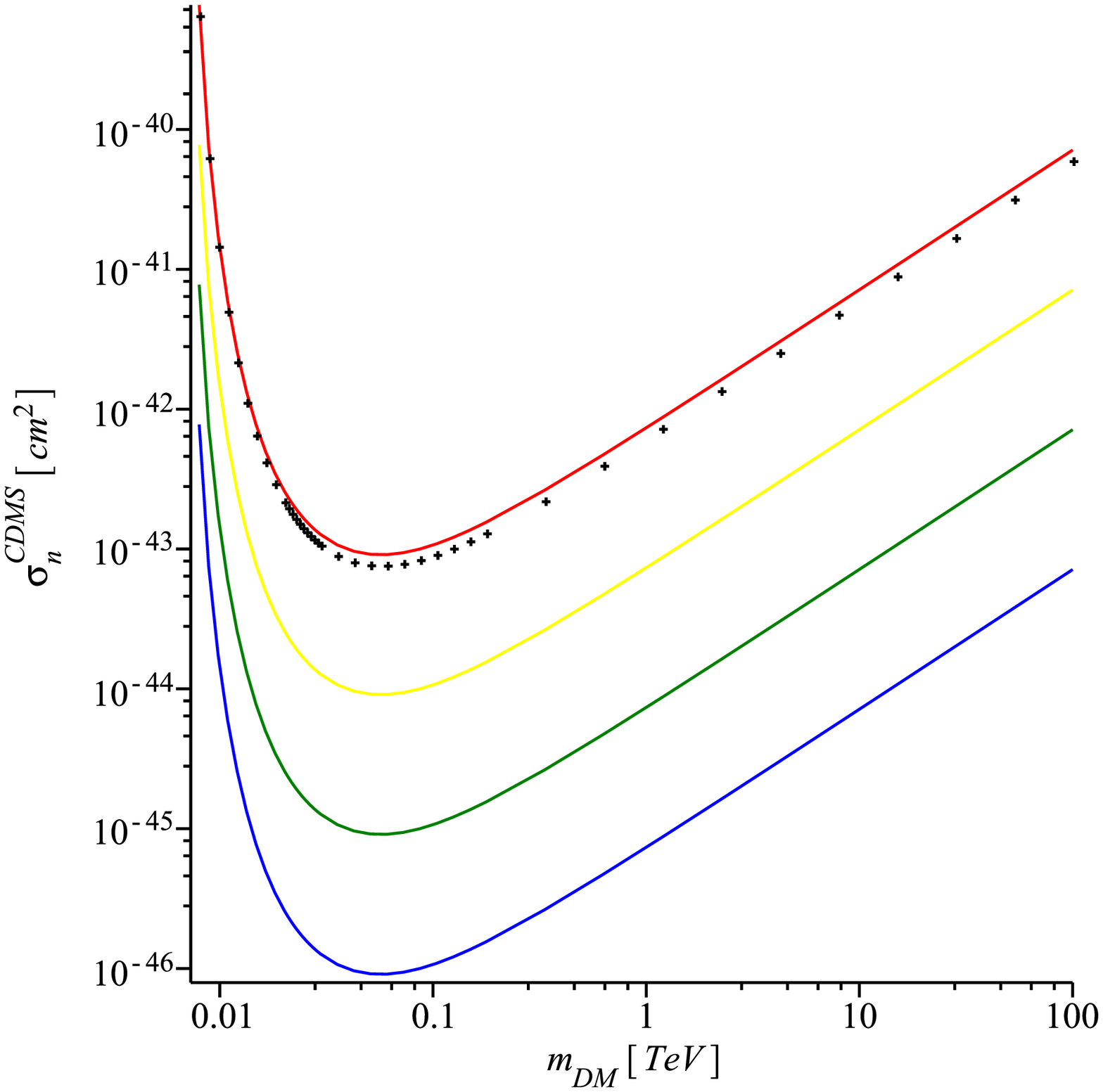}
\end{center}
\end{minipage}
\hspace{0cm}
\begin{minipage}[b]{0.5\linewidth}
\begin{center}
\includegraphics[scale=0.3]{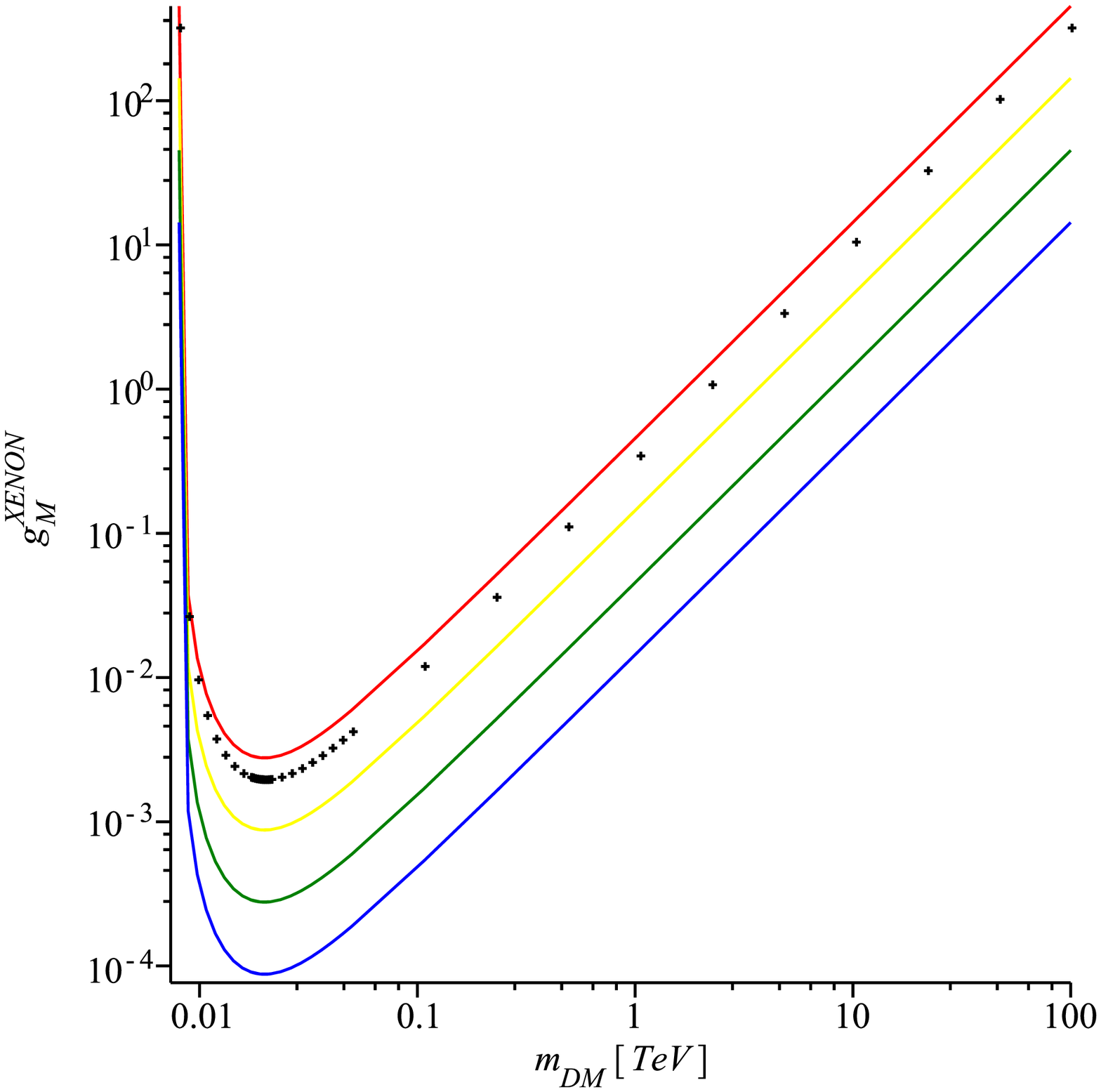}
\includegraphics[scale=0.3]{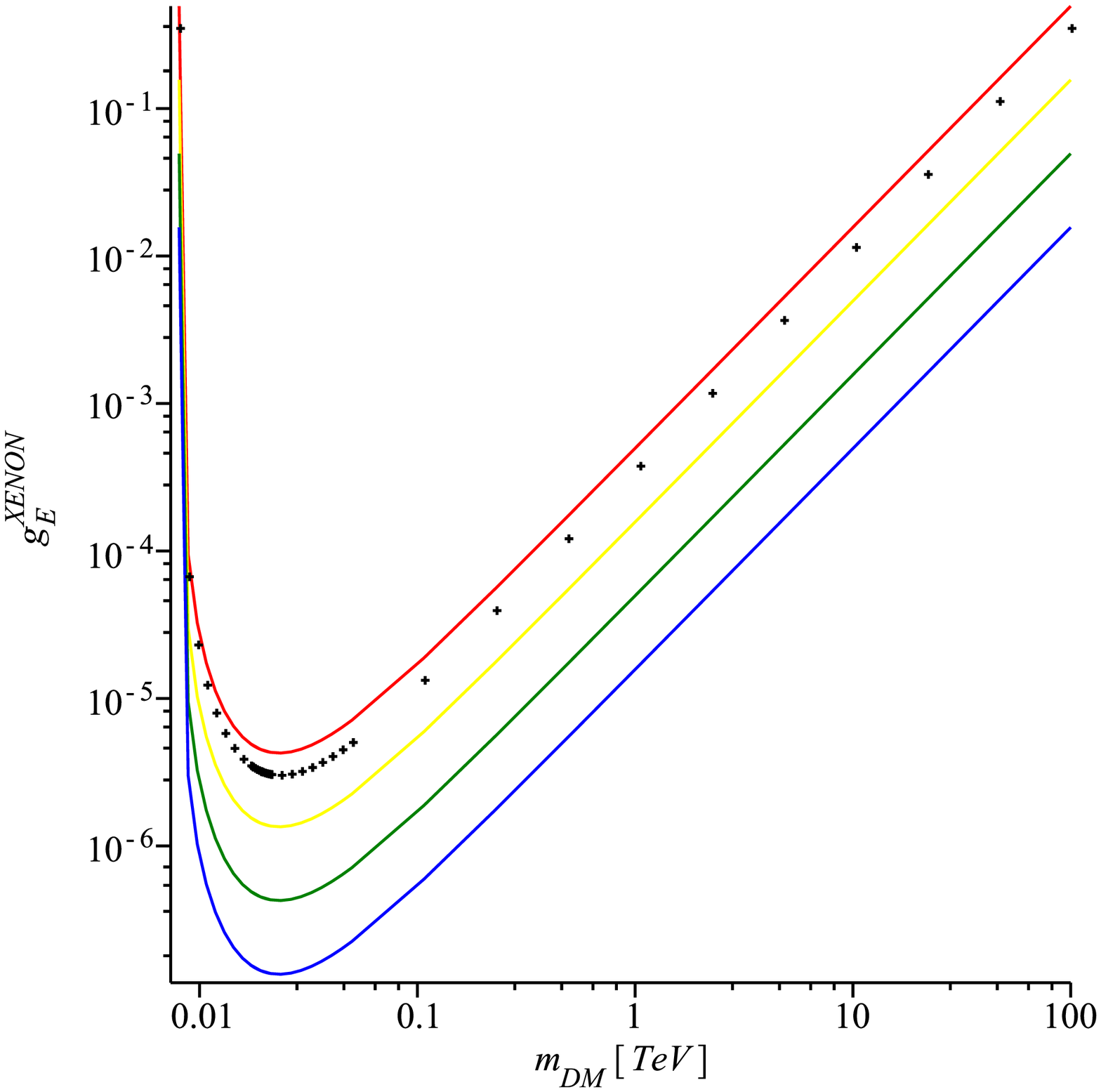}
\includegraphics[scale=0.3]{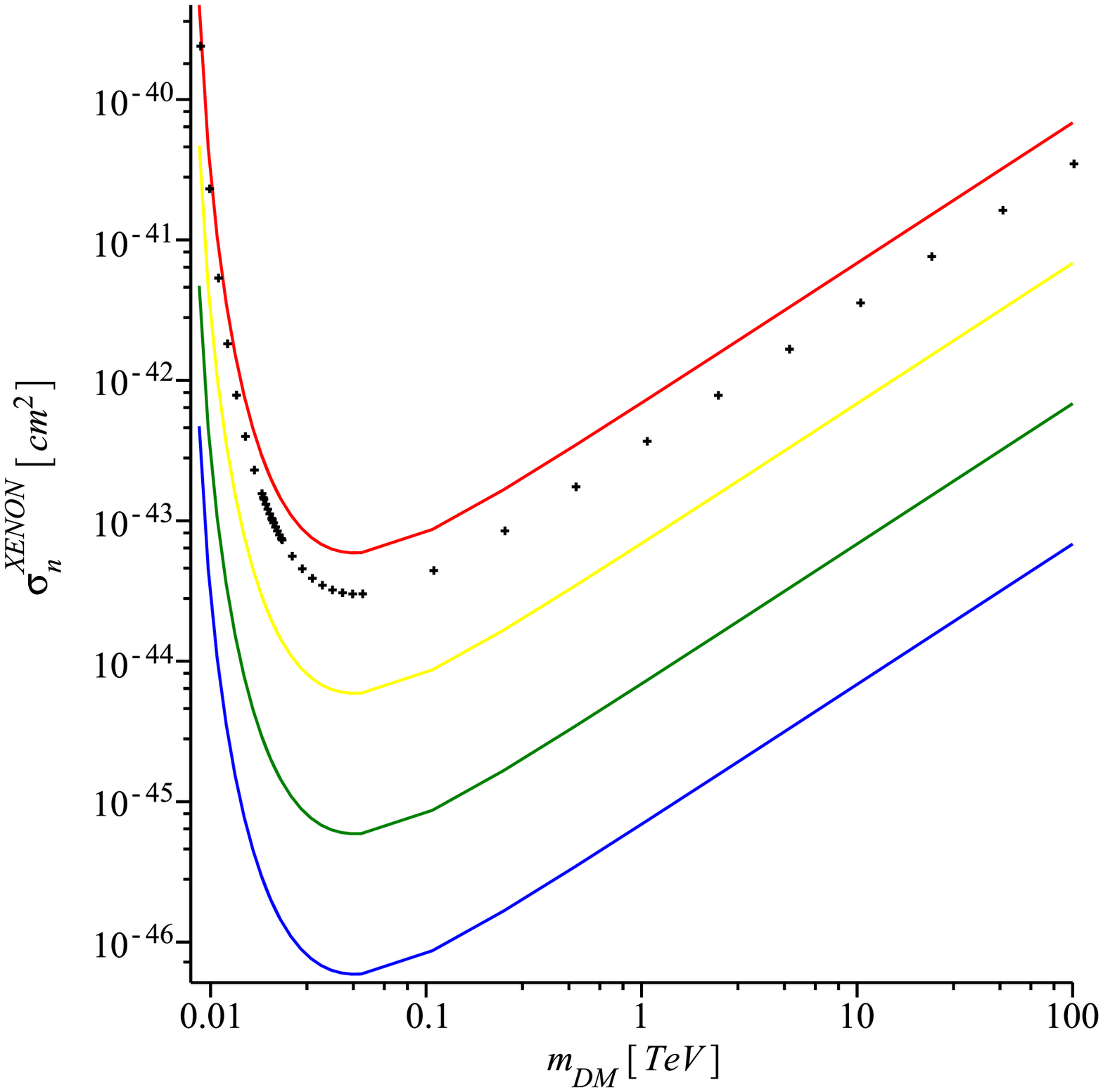}
\end{center}
\end{minipage}
\caption{Fermionic DM contour plots relevant to CDMS (left) and XENON (right).  The magnetic dipole moment, the electric dipole moment and the constant contact interaction are shown in the top, middle and bottom plots respectively.  The red, yellow, green and blue curves correspond to $0.1$, $1$, $10$ and $100$ ${\rm event}^{-1}\cdot{\rm kg}\cdot{\rm year}$ respectively.  The black crosses correspond to the $90\%$ confidence level limits for the CDMS and XENON data.}
\label{fig:ContourPlot}
\end{figure}

A partial list of the direct detection experiments includes : CDMS \cite{Ahmed:2009zw}, CoGeNT \cite{Aalseth:2010vx}, EDELWEISS \cite{Armengaud:2009hc}, IGEX \cite{Morales:2000ns}, HDMS \cite{KlapdorKleingrothaus:2002pg,Bednyakov:2008zza}, XENON \cite{Aprile:2010um}, ZEPLIN \cite{Lopes:2010zz}, DAMA/LIBRA \cite{Bernabei:2010mq}, NAIAD \cite{Alner:2005kt}, KIMS \cite{Myung:2009zz}, PICASSO \cite{Beltran:2010zz,Piro:2010qd}, KAMIOKA \cite{Miuchi:2002zp}, NEWAGE \cite{Miuchi:2010hn}, SIMPLE \cite{Felizardo:2010mi}, CRESST \cite{Angloher:2008zz,Roth:2010zza}, WARP \cite{Benetti:2007cd} and ORPHEUS \cite{Borer:2004qf}.  The relevant features of each experiment are summarized in table \ref{tab:Exp}.

A list of planned experiments includes : GENIUS (Ge) \cite{Tomei:2003vc}, XMASS (LXe) \cite{Sekiya:2010bf}, LUX (LXe) \cite{McKinsey:2010zz}, ANAIS (NaI(Tl)) \cite{Amare:2006pe}, COUPP (CF$_3$Br) \cite{Collar:2007xn}, EURECA (CaWO$_4$ \& Ge) \cite{Kraus:2008zz}, ArDM (LAr) \cite{Haranczyk:2010cf}, DEAP/CLEAN (LAr) \cite{Giuliani:2010zz} and ROSEBUD (Al \& O) \cite{Amare:2006ph}.

The DM direct search experiments we will focus on are CDMS \cite{Ahmed:2009zw}, XENON \cite{Aprile:2010um} and DAMA \cite{Bernabei:2010mq}.  Notice that table \ref{tab:Elements} already predicts larger magnetic dipole moment differential scattering rates per unit detector mass for experiments with iodine than for experiments with xenon.  We therefore expect a larger rate at DAMA than at XENON in the magnetic dipole moment case, which could in principle lead to an unified explanation of the signals, a possibility that is ruled out for WIMPs with the usual constant contact interactions.

\subsection{CDMS and XENON}\label{subsec:CDMSXENON}

For CDMS (Ge) and XENON (Xe), the allowed regions for the magnetic dipole moment, the electric dipole moment, and the constant contact interaction are shown in figure \ref{fig:ContourPlot} as well as relevant contours.  Exposures of $0.1$, $1$, $10$ and $100$ ${\rm event}^{-1}\cdot{\rm kg}\cdot{\rm year}$ are depicted by the red, yellow, green and blue curves respectively.  The $90\%$ confidence level limits for CDMS and XENON are represented by the black crosses.

From figure \ref{fig:ContourPlot} one notices that order one magnetic Land\'{e} factors $g_M\sim1$ and thus strongly-coupled dark sectors are excluded by CDMS and XENON for fermionic DM masses $m_{\rm DM}\lesssim1$ TeV.  A similar statement for the electric Land\'{e} factor $g_E$ cannot be made since $(\Lambda_{\rm CP}/\Lambda_{\rm DM})^2g_E$ is model-dependent.

\subsection{DAMA}\label{subsec:DAMA}

\begin{figure}[p]
\begin{minipage}[b]{0.5\linewidth}
\begin{center}
\includegraphics[scale=0.35]{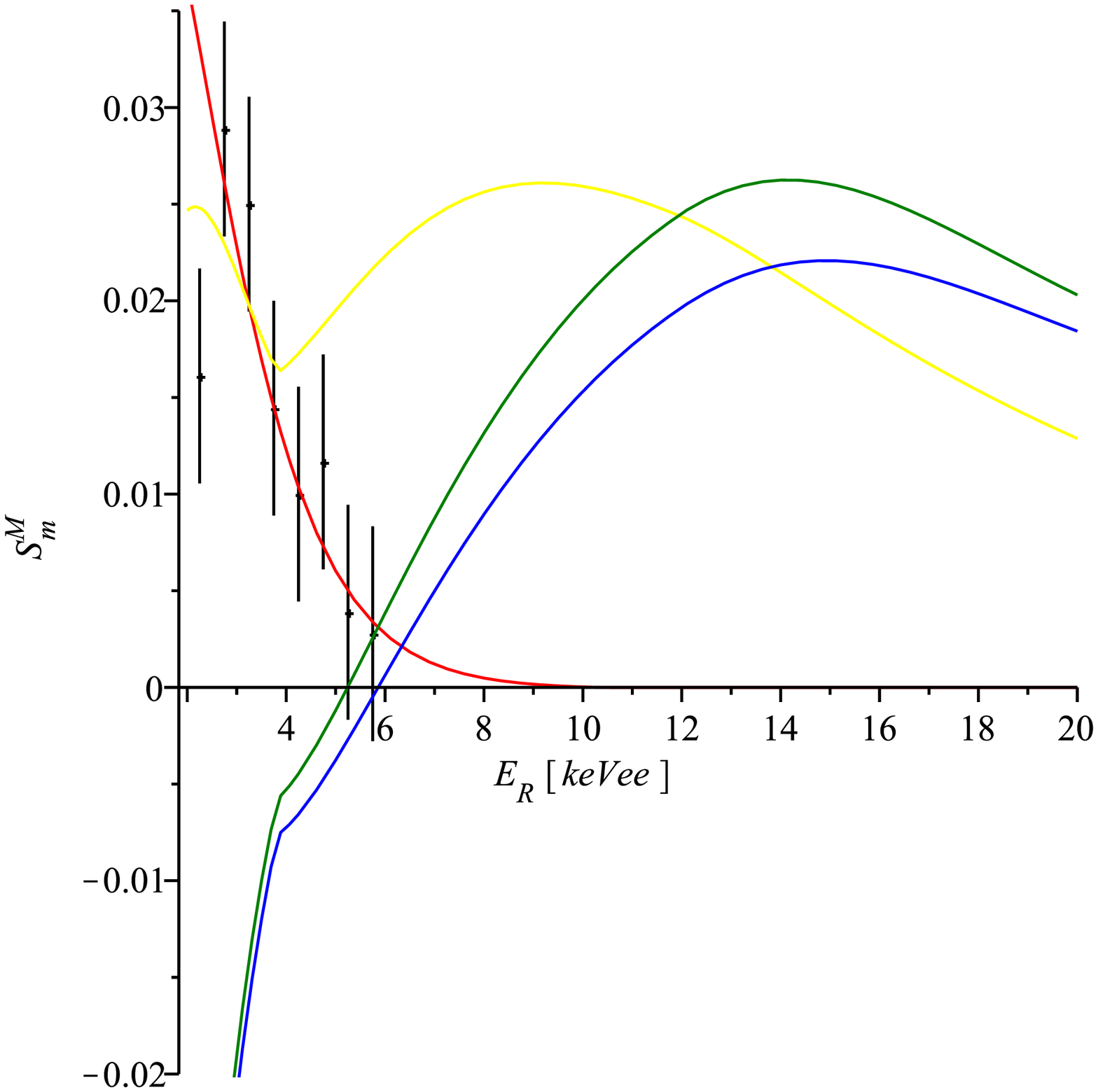}
\end{center}
\end{minipage}
\hspace{0cm}
\begin{minipage}[b]{0.5\linewidth}
\begin{center}
\includegraphics[scale=0.35]{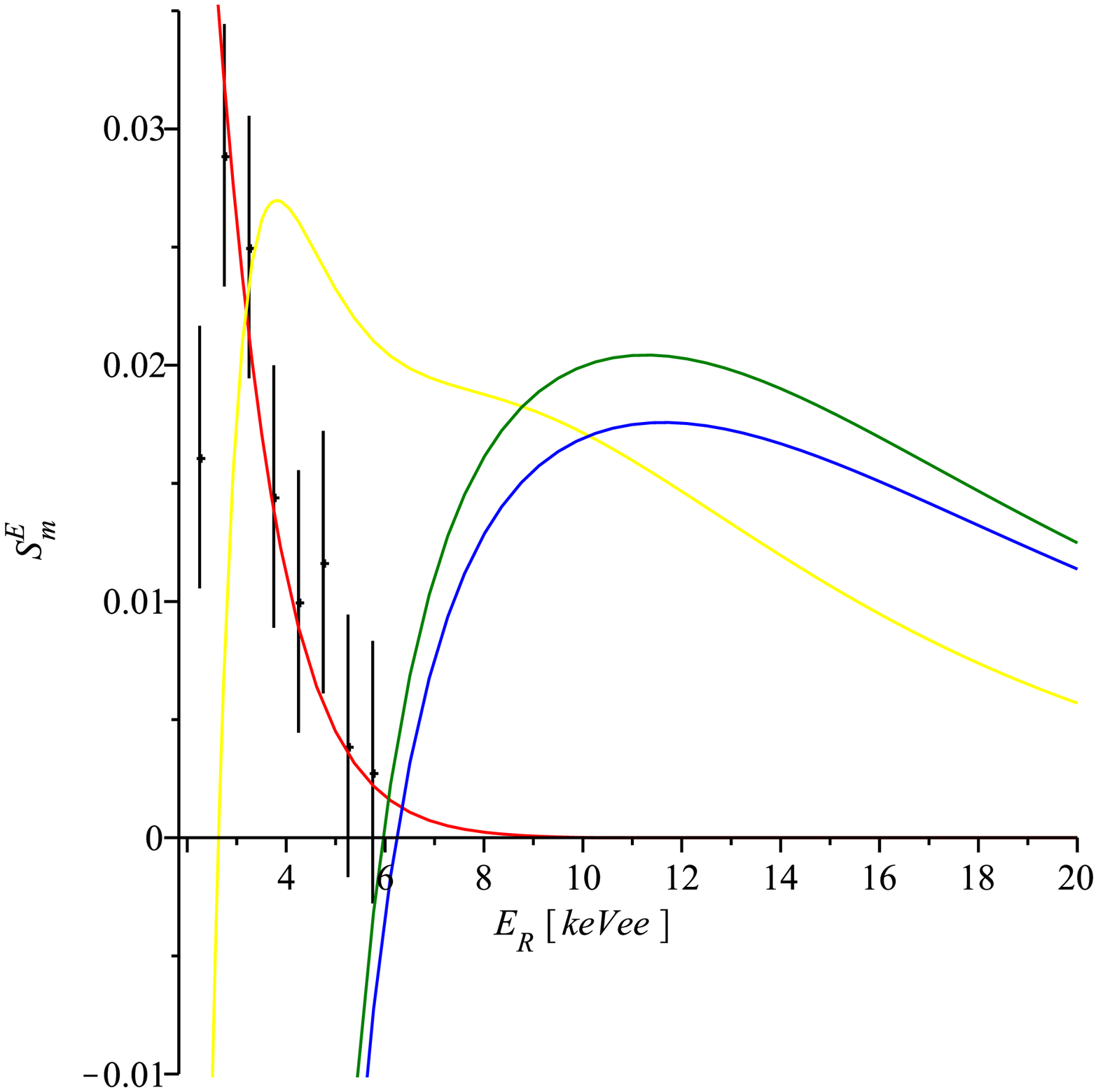}
\end{center}
\end{minipage}
\begin{center}
\includegraphics[scale=0.35]{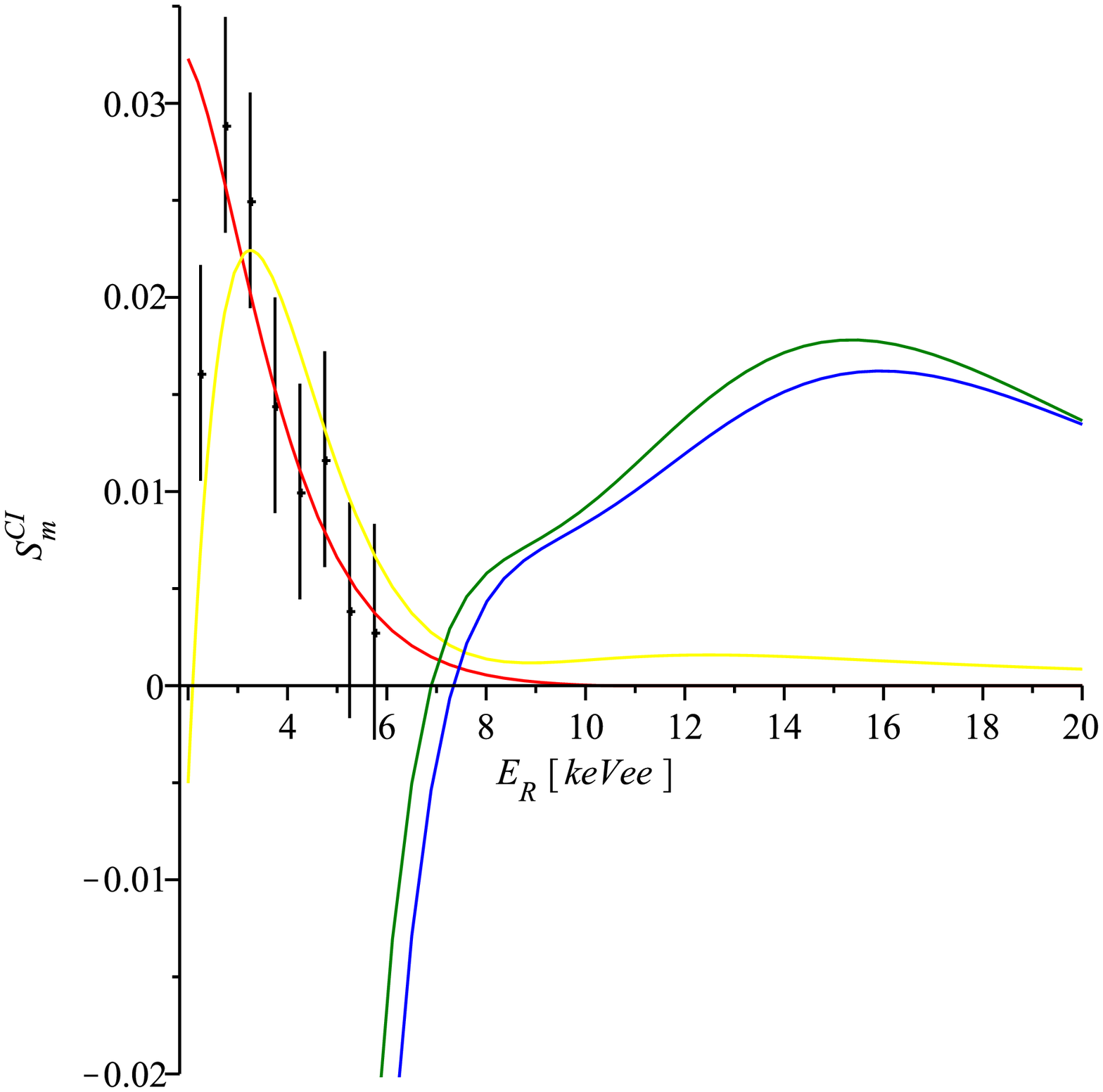}
\end{center}
\caption{Modulation amplitude $S_m$ in ${\rm event}\cdot{\rm kg}^{-1}\cdot{\rm day}^{-1}\cdot{\rm keVee}^{-1}$ relevant to DAMA for fermionic DM magnetic dipole moment (top left), electric dipole moment (top right) and constant contact interaction (bottom middle).  The red, yellow, green and blue curves correspond to fermionic DM masses and gyromagnetic ratios ($m_{\rm DM}$, $g_M$) of ($0.01$ TeV, $0.02$), ($0.1$ TeV, $0.6$), ($1$ TeV, $20$) and ($10$ TeV, $600$) respectively, fermionic DM masses and gyromagnetic ratios ($m_{\rm DM}$, $g_E$) of ($0.01$ TeV, $6\cdot10^{-5}$), ($0.1$ TeV, $2.5\cdot10^{-3}$), ($1$ TeV, $0.1$) and ($10$ TeV, $3$) respectively and fermionic DM masses and total cross section per nucleon ($m_{\rm DM}$, $\sigma_n$) of ($0.01$ TeV, $8\cdot10^{-40}$ cm$^2$), ($0.1$ TeV, $4\cdot10^{-40}$ cm$^2$), ($1$ TeV, $4\cdot10^{-38}$ cm$^2$) and ($10$ TeV, $4\cdot10^{-37}$ cm$^2$) respectively.  The black crosses with error bars correspond to the DAMA data consistent with non-vanishing modulation amplitude.  The gyromagnetic ratios and the total cross sections per nucleon are chosen such that the modulation amplitudes are of the same order of magnitude than the DAMA data.}
\label{fig:ModulationDAMA}
\end{figure}

For DAMA (NaI), the modulation amplitude $S_m$ for the magnetic dipole moment, the electric dipole moment and the constant contact interaction are shown in figure \ref{fig:ModulationDAMA} for different fermionic DM masses.  The quenching factors used for sodium and iodine are $0.3$ and $0.09$ respectively (the effect of thallium is negligible due to its low concentration).  Fermionic DM masses of $0.01$, $0.1$, $1$ and $10$ TeV are represented by the red, yellow, green and blue curves.  DAMA data consistent with non-vanishing modulation amplitude are shown by the black crosses with error bars.  The gyromagnetic ratios and the total cross section per nucleon are chosen such that the modulation amplitudes and the DAMA data are of the same order of magnitude.

With the help of figure \ref{fig:ModulationDAMA} it is obvious that DAMA can be explained by a light DM with mass of order $m_{\rm DM}\sim10$ GeV.  In the case of magnetic dipole moment interactions, the gyromagnetic ratio must be $g_M\sim0.02$.  For electric dipole moment interactions, the gyromagnetic ratio must be $g_E\sim6\cdot10^{-5}$.  Finally, for s-wave scattering, the total cross section per nucleon must be $\sigma_n\sim8\cdot10^{-40}$ cm$^2$.  It is interesting to note that the total differential rate in function of the recoil energy is consistent with DAMA only for DM with mass of order $m_{\rm DM}\sim10$ GeV.  For DM with masses $0.1$, $1$ and $10$ TeV, the gyromagnetic ratios and the total cross section per nucleon needed to obtain modulation amplitudes of the same order of magnitude than DAMA lead to total differential rates inconsistent with DAMA.

%%%%%%%%%%%%%%%%%%%%%%%%%%%%%%%%%%%%%%%%%%%%%%%%%%%%%%%%%%%%%%%%%%%%%%%%%%%%%%%%%%%%%%%%%%%%%%%%%%%%%%%%%%%%%%%%%%%%%%%%%%%%%%%%%%%%%%%%%%%%%%%%%%%%

\section{Discussion and Conclusion}\label{sec:conc}

Fermionic DM with magnetic or electric dipole moments can explain CDMS/XENON and DAMA with a fermionic DM of mass $m_{\rm DM}\sim10$ GeV and gyromagnetic ratios $g_M\sim0.02$ or $g_E\sim6\cdot10^{-5}$ respectively.  This is not the case for a typical WIMP with constant contact interactions.  Indeed, to explain DAMA, s-wave scatterings with total cross section per nucleon $\sigma_n\sim8\cdot10^{-40}$ cm$^2$ is needed, which is excluded both by CDMS and XENON.  The agreement between CDMS/XENON and DAMA for fermionic DM with magnetic or electric dipole moments is very constrained and can be destroyed with more statistics.

If DAMA is discarded, CDMS and XENON alone allow a wide range of fermionic DM masses (few GeV to a few tens of TeV) with gyromagnetic ratios encompassing several orders of magnitude.  Order one magnetic Land\'{e} factors are not excluded for fermionic DM masses $m_{\rm DM}\gtrsim1$ TeV and thus strongly-coupled dark sectors with scales $\Lambda_{\rm DM}\gtrsim1$ TeV are possible.  Strongly-coupled dark sectors, a common feature of several direct mediation and technicolor models with dark baryons as fermionic DM, thus cannot explain CDMS/XENON and DAMA.  However, fermionic DM with dipole moments can explain all experiments for a DM mass $m_{\rm DM}\sim10$ GeV without relying on any subtleties.

From table \ref{tab:Elements}, interesting experiments which could shed light on the nature of the DM candidate from the point of view of magnetic dipole moments, are experiments with cesium or fluorine, like KIMS (cesium) or PICASSO, KAMIOKA and NEWAGE (fluorine), since these can lead to more stringent constraints for the magnetic dipole moment operators.

After this work was substantially completed, we noticed two papers discussing DM magnetic interactions with nuclei in the context of inelastic DM \cite{Chang:2010en} and with respect to CoGeNT \cite{Barger:2010gv}.  The latter work on CoGeNT is consistent with our conclusions.

\subsection*{Acknowledgments}

We would like to thank Michael Dine, Andreas Fuhrer, Richard Gaitskell, Aneesh Manohar, Dan McKinsey, Joel Primack, Richard Schnee, Neal Weiner, and especially Stephano Profumo for valuable discussions and comments on this manuscript.  The research of TB was supported in part by DE-FG03-92ER40689.  The research of JFF was supported in part by DOE grant DOE-FG03-97ER40546.  The research of ST was supported in part by DOE grant DE-FG02-96ER40949.

%%%%%%%%%%%%%%%%%%%%%%%%%%%%%%%%%%%%%%%%%%%%%%%%%%%%%%%%%%%%%%%%%%%%%%%%%%%%%%%%%%%%%%%%%%%%%%%%%%%%%%%%%%%%%%%%%%%%%%%%%%%%%%%%%%%%%%%%%%%%%%%%%%%%

\appendix

\section{Effective Lagrangian for Fermionic Dark Matter}\label{app:EffLag}

For definiteness, we summarize in this appendix the results of \cite{Bagnasco:1993st} using the heavy field formalism \cite{Georgi:1990um} for nuclei of arbitrary nuclear spins.

For a fermionic DM, the most relevant operator is the magnetic dipole moment which is the only dimension $5$ operator.  A fermionic DM can also have an electric dipole moment which, in realistic models, is a dimension $6$ operator due to the intrinsic P and T violation of the operator.  The effective Lagrangian of the magnetic and electric dipole moments for a fermionic DM is thus
\begin{equation}\label{eqn:LagDM}
\delta\mathcal{L}_{\rm DM}=\bar{\psi}(i\gamma^\mu\partial_\mu-m_{\rm DM})\psi+\frac{g_Me}{8m_{\rm DM}}\bar{\psi}\sigma^{\mu\nu}\psi F_{\mu\nu}+\frac{g_Ee}{8m_{\rm DM}}\bar{\psi}\sigma^{\mu\nu}\psi \widetilde{F}_{\mu\nu}
\end{equation}
where $m_{\rm DM}$ is the fermionic DM mass and $g_M\sim4m_{\rm DM}/\Lambda_{\rm DM}$ and $g_E\sim4m_{\rm DM}\Lambda_{\rm DM}/\Lambda_{\rm CP}^2$ are the DM equivalent of the Land\'{e} factors.  The scales relevant for the dipole moment operators are $\Lambda_{\rm DM}$ which is the dark sector scale and $\Lambda_{\rm CP}>\Lambda_{\rm DM}$ which is the scale associated with CP violation \cite{Bagnasco:1993st}.  The gyromagnetic ratios $g_M$ and $(\Lambda_{\rm CP}/\Lambda_{\rm DM})^2g_E$ should be order one numbers in strongly-coupled theories and loop-suppressed numbers in weakly-coupled theories.  This Lagrangian leads to a magnetic dipole moment $\mu_{\rm DM}=g_Me/4m_{\rm DM}$ and an electric dipole moment $d_{\rm DM}=g_Ee/4m_{\rm DM}$ for the fermionic DM.  The normalization is consistent with the electron magnetic dipole moment\footnote{The normalization of \cite{Cho:2010br} is not consistent with the usual definitions of the Land\'e $g$-factor and the magnetic dipole moment.} where $\mu_e\equiv|\boldsymbol{\mu}_e|=g_ee|\boldsymbol{S}|/2m_e=g_ee/4m_e$ with $\boldsymbol{S}$ the spin operator and $g_e=2$ the correct gyromagnetic ratio.

For a nucleus of arbitrary nuclear spin $J$ and atomic and mass numbers $Z$ and $A$, the heavy field formalism \cite{Georgi:1990um} can be used in the non-relativistic limit and leads to the following effective Lagrangian in the rest frame of the nucleus,
\begin{equation}\label{eqn:LagN}
\delta\mathcal{L}_N=N^\dagger\left(iD_0+\frac{\boldsymbol{D}\cdot\boldsymbol{D}}{2m_N}-\boldsymbol{\mu}_N\cdot\boldsymbol{B}\right)N.
\end{equation}
Here $m_N$ is the nucleus mass, $D_\mu=\partial_\mu-iZeA_\mu$ is the covariant derivative, $\mu_N\equiv|\boldsymbol{\mu}_N|\propto|\boldsymbol{S}|$ is the nuclear magnetic moment and $\boldsymbol{B}$ is the magnetic field.

In the non-relativistic limit, the differential cross sections are easily obtained from the above Lagrangians $\delta\mathcal{L}=\delta\mathcal{L}_{\rm DM}+\delta\mathcal{L}_N$, the nuclear magnetic moment $\mu_N/\mu_n$ in units of the nuclear Bohr magneton $\mu_n=e/2m_n$, the nuclear spin form factor $F_s(E_R)$, the form factor which accounts for the loss of coherence over the nucleus at finite momentum transfer $F_c(E_R)$ ($|F_c(0)|=1$), and the fermionic DM velocity in the lab frame $v$.  Since the fermionic DM magnetic moment couples to the nuclear magnetic moment and coherently to the nuclear charge current in the fermionic DM rest frame, the differential cross section in the non-relativistic limit for the magnetic dipole moment contains two terms, the spin-dependent (SD) and the spin-independent (SI) terms\footnote{The result of \cite{Cho:2010br} for the magnetic dipole moment is not divided into well-defined SI and SD contributions.  Indeed the contribution coming from the vector current of the nucleus ($\bar{\psi}\gamma^\mu\psi$) contains both a SI piece and a SD piece as can be easily seen from the Gordon identity.}.  In the lab frame the differential cross sections in the non-relativistic limit are thus
\begin{eqnarray}
\frac{d\sigma_M}{dE_R} &=& \frac{\pi(g_M\alpha)^2}{4(m_{\rm DM}+m_N)^2E_R^{\rm max}}\left\{\frac{2(J+1)}{3J}\left(\frac{\mu_NA}{\mu_n}\right)^2|F_s(E_R)|^2\right.\nonumber\\
 && \hspace{1cm}\left.+Z^2\left(\frac{(m_{\rm DM}+m_N)^2}{m_{\rm DM}^2}\frac{E_R^{\rm max}}{E_R}-\frac{2m_N}{m_{\rm DM}}-1\right)|F_c(E_R)|^2\right\}\label{eqn:CrossSectionM}\\
\frac{d\sigma_E}{d E_R} &=& \frac{\pi(g_EZ\alpha)^2}{2(m_{\rm DM}+m_N)^2E_R^{\rm max}}\frac{m_N}{E_R}|F_c(E_R)|^2\label{eqn:CrossSectionE}
\end{eqnarray}
where $E_R$ is the nuclear recoil kinetic energy in the lab frame and $E_R^{\rm max}=2m_Nm_{\rm DM}^2v^2/(m_{\rm DM}+m_N)^2$ is the maximum nuclear recoil energy, $E_R\leq E_R^{\rm max}$.  Notice that the coherent contribution to the magnetic dipole moment differential cross section (SI) and the electric dipole moment differential cross section have an IR enhancement at low recoil energy.  Moreover, in order not to introduce an extra scale, the dark sector is usually assumed not to violate CP and the electric dipole moment is forgotten altogether.  However, since the electric dipole moment differential cross section is enhanced by a factor $\sim m_N/E_R\sim10^{6-8}$ compared to the magnetic dipole moment differential cross section, the electric dipole moment could in principle be the dominant effect and thus should not be set aside.

It should be emphasized that the differential cross section in the non-relativistic limit for a constant contact interaction (i.e. s-wave nucleon scattering) is different than that for the dipole moments.  Indeed, in the non-relativistic limit, the differential cross section for a constant contact interaction is
\begin{equation}\label{eqn:CrossSectionCI}
\frac{d\sigma_{CI}}{d E_R}=\frac{A^2\sigma_n}{E_R^{\rm max}}\frac{\widetilde{m}_{{\rm DM},N}^2}{\widetilde{m}_{{\rm DM},n}^2}|F_c(E_R)|^2
\end{equation}
where $\sigma_n\equiv\sigma({\rm DM}+n\rightarrow{\rm DM}+n)$ is the total cross section per nucleon and $\widetilde{m}_{{\rm DM},N}$ and $\widetilde{m}_{{\rm DM},n}$ are the reduced masses of the fermionic DM-nucleus system and fermionic DM-nucleon system respectively.  A comparison between the differential cross sections for a constant contact interaction and for the dipole moments would eventually allow distinguishing between s-wave and dipole scatterings.

Notice that the nuclear spin contribution to the magnetic dipole moment differential cross section (SD) and the contact interaction differential cross section increase with the nuclear mass number $A$ while the coherent contribution to the magnetic dipole moment differential cross section (SI) and the electric dipole moment differential cross section increase with the nuclear atomic number $Z$.  Therefore the dominant contribution to the differential cross section depends strongly on the experiment.

The distributions of nuclear `scattering centres' and spin within the nucleus are well approximated by
\begin{eqnarray}\label{eqn:FcFs}
|F_c(E_R)|^2 &=& 9\left[\frac{\sin(qR_c)-qR_c\cos(qR_c)}{(qR_c)^3}\right]^2e^{-(qs)^2}\\
|F_s(E_R)|^2 &=& \left\{\begin{array}{lcl}\left[\frac{\sin(qR_s)}{qR_s}\right]^2 & \mbox{for} & qR_s<2.55,\,qR_s>4.5\\
0.047 & \mbox{for} & 2.55\leq qR_s\leq4.5\end{array}\right.
\end{eqnarray}
with momentum transfer $q=\sqrt{2m_NE_R}$, effective `scattering centres' and spin nuclear radii $R_c=1.14\,A^{1/3}$ fm and $R_s=1.0\,A^{1/3}$ fm respectively and nuclear skin thickness $s=0.9$ fm \cite{Lewin:1995rx}.

\begin{figure}[t]
\begin{minipage}[b]{0.5\linewidth}
\begin{center}
\includegraphics[scale=0.35]{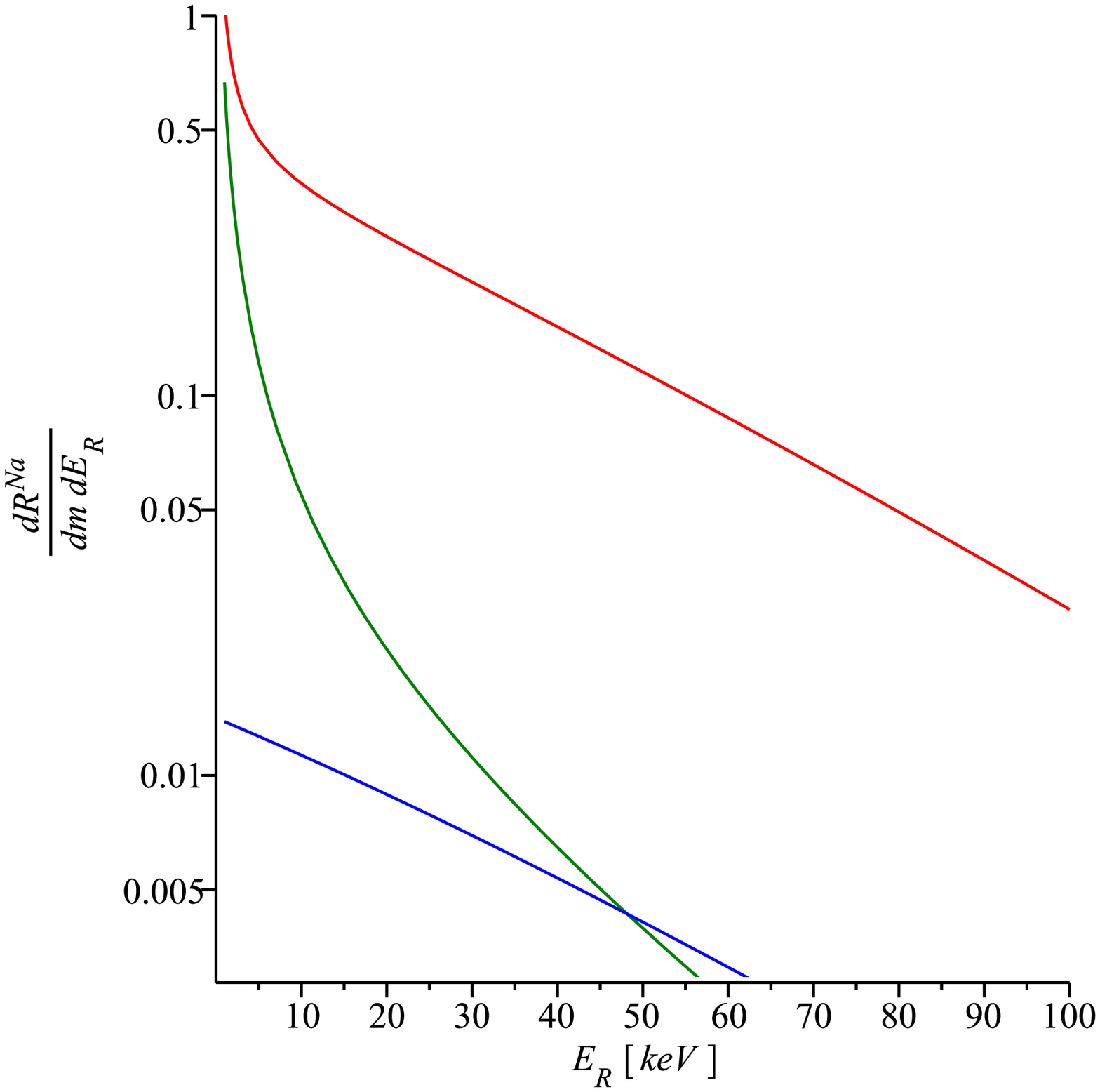}
\end{center}
\end{minipage}
\hspace{0cm}
\begin{minipage}[b]{0.5\linewidth}
\begin{center}
\includegraphics[scale=0.35]{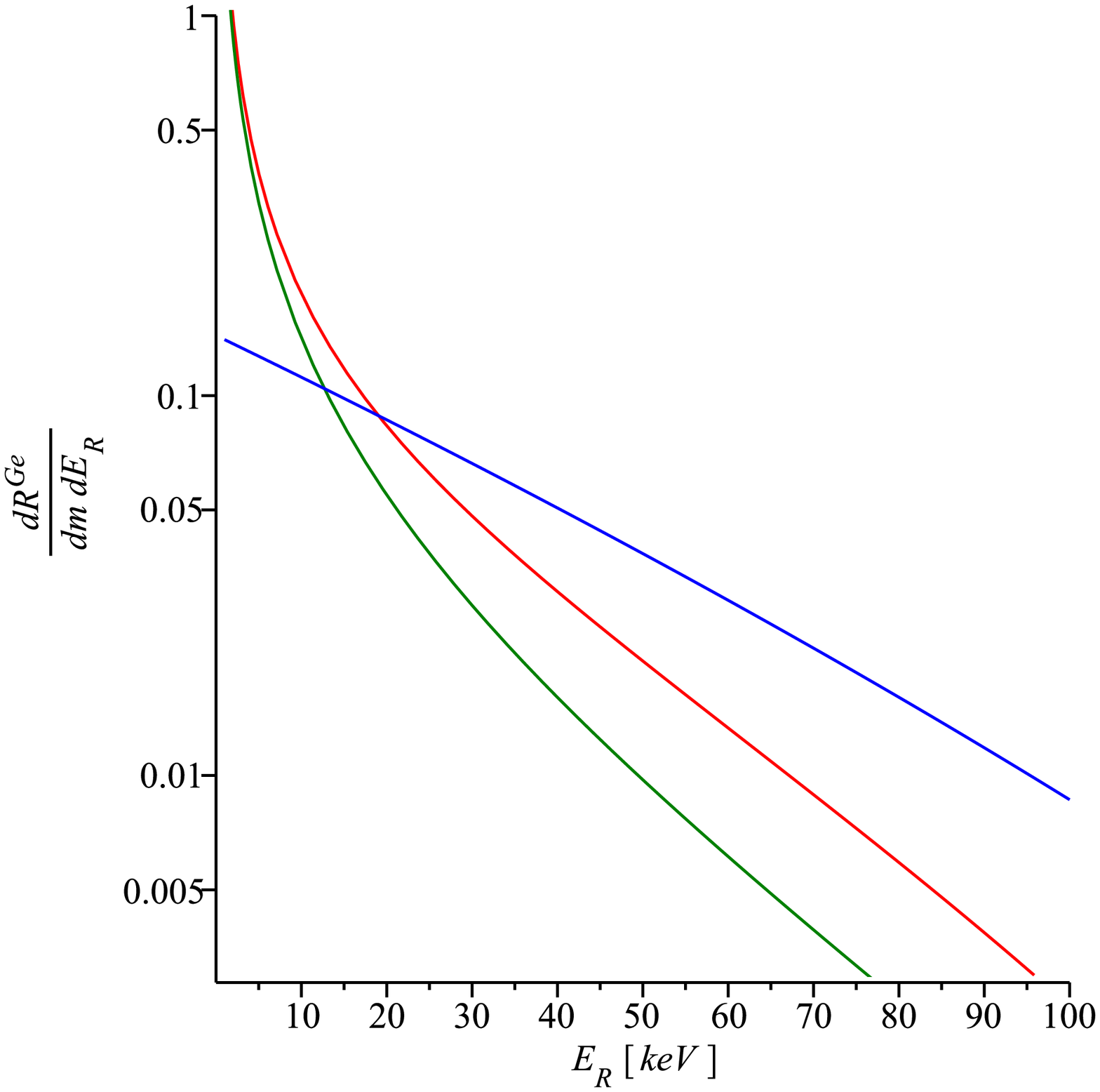}
\end{center}
\end{minipage}
\caption{Differential scattering rates as a function of the nuclear recoil energy for $^{23}$Na (left) and $^{73}$Ge (right).  The red, green and blue curves correspond to the magnetic dipole moment scattering rate, the electric dipole moment scattering rate and the constant contact interaction scattering rate in ${\rm event}\cdot{\rm kg}^{-1}\cdot{\rm day}^{-1}\cdot{\rm keV}^{-1}$ respectively.  The gyromagnetic ratios are chosen to be $g_M=4$ and $g_E=4\cdot10^{-3}$, the total cross section per nucleon is assumed to be $\sigma_n=1\cdot10^{-40}$ ${\rm cm}^2$ and the fermionic DM mass is assumed to be $m_{\rm DM}=1$ TeV.}\label{fig:Rate}
\end{figure}

Finally, to obtain the number of events seen for each experiment, one needs the differential scattering rate per unit detector mass $m$, which is given by
\begin{equation}\label{eqn:DiffScatRate}
\frac{dR}{dm\,dE_R}=\frac{\rho}{m_Nm_{\rm DM}}\left\langle\frac{d\sigma}{dE_R}v\right\rangle
\end{equation}
where $\rho=300\,{\rm TeV}\cdot{\rm m}^{-3}$ is the fermionic DM density and
\begin{equation}
\left\langle\frac{d\sigma}{dE_R}v\right\rangle=\int_{v_{\rm min}(E_R)}^{v_{\rm max}}dv f(v)v\frac{d\sigma}{dE_R}(v,E_R)
\end{equation}
is the velocity-averaged differential cross section weighted by the fermionic DM speed $v$.  Here $f(v)$ is the fermionic DM speed distribution in the lab frame, $v_{\rm min}=(1+m_N/m_{\rm DM})\sqrt{E_R/2m_N}$ is the minimum fermionic DM speed in the lab frame and $v_{\rm max}$ is the maximum fermionic DM speed in the lab frame.  The fermionic DM velocity distribution is assumed to be a (canonically normalized) Maxwell-Boltzmann distribution in the galactic rest frame,
\begin{equation}
f_{\rm galaxy}(\boldsymbol{v})=\frac{1}{\pi^{3/2}\bar{v}^3}e^{-\boldsymbol{v}^2/\bar{v}^2}
\end{equation}
where $\bar{v}=230\,{\rm km}\cdot{\rm s}^{-1}$ is the fermionic DM most probable speed in the galactic rest frame \cite{Lewin:1995rx}.  After a Galilean transformation, the fermionic DM speed distribution in the lab frame is
\begin{equation}\label{eqn:DMfv}
f(v)=\left\{\begin{array}{lcl}\frac{v}{\pi^{1/2}\bar{v}u}\left(e^{-(v-u)^2/\bar{v}^2}-e^{-(v+u)^2/\bar{v}^2}\right) & \mbox{for} & 0\leq v\leq v_{\rm esc}-u\\
\frac{v}{\pi^{1/2}\bar{v}u}\left(e^{-(v-u)^2/\bar{v}^2}-e^{-v_{\rm esc}^2/\bar{v}^2}\right) & \mbox{for} & v_{\rm esc}-u<v\leq v_{\rm esc}+u\end{array}\right.
\end{equation}
where $u\equiv u(y)=[244+15\sin(2\pi y)]\,{\rm km}\cdot{\rm s}^{-1}$ is the earth's speed in the galactic rest frame with $y$ the time elapsed from March 2nd in years \cite{Lewin:1995rx} and $v_{\rm esc}=600\,{\rm km}\cdot{\rm s}^{-1}$ is the local galactic escape velocity in the galactic rest frame \cite{Lewin:1995rx}.

As an example, the differential scattering rates for fermionic DM with magnetic dipole moment, electric dipole moment and constant contact interaction are shown in figure \ref{fig:Rate} for two different elements, $^{23}$Na and $^{73}$Ge.  Notice that the magnetic and electric dipole moment differential scattering rates are similar at low nuclear recoil energies due to the IR enhancement.

%%%%%%%%%%%%%%%%%%%%%%%%%%%%%%%%%%%%%%%%%%%%%%%%%%%%%%%%%%%%%%%%%%%%%%%%%%%%%%%%%%%%%%%%%%%%%%%%%%%%%%%%%%%%%%%%%%%%%%%%%%%%%%%%%%%%%%%%%%%%%%%%%%%%

%%%%%%%%%%%%%%%%%%%%%%%%%%%%%%%%%%%%%%%%%%%%%%%%%%%%%%%%%%%%%%%%%%%%%%%%%%%%%%%%%%%%%%%%%%%%%%%%%%%%%%%%%%%%%%%%%%%%%%%%%%%%%%%%%%%%%%%%%%%%%%%%%%%%
%%%%%%%%%%%%%%%%%%%%%%%%%%%%%%%%%%%%%%%%%%%%%%%%%%%%%%%%%%%%%%%%%%%%%%%%%%%%%%%%%%%%%%%%%%%%%%%%%%%%%%%%%%%%%%%%%%%%%%%%%%%%%%%%%%%%%%%%%%%%%%%%%%%%

\end{document}